\documentclass[aps,pre,twocolumn,amsmath,amssymb,superscriptaddress,floatfix]{revtex4} 

\usepackage{graphicx}
\usepackage{bm}
\usepackage[usenames]{color}
\usepackage{amsmath}
\bibstyle{apsrev.bib}

\newcommand{\be}{\begin{equation}}
\newcommand{\ee}{\end{equation}}
\newcommand{\beqn}{\begin{eqnarray}}
\newcommand{\eeqn}{\end{eqnarray}}

\begin{document}

\title{Entanglement between random and clean quantum spin chains}
\author{R\'obert Juh\'asz}
\email{juhasz.robert@wigner.mta.hu}
\affiliation{Wigner Research Centre for Physics, Institute for Solid State
Physics and Optics, H-1525 Budapest, P.O.Box 49, Hungary}
\author{Istv\'an A. Kov\'acs}
\email{kovacs.istvan@wigner.mta.hu}
\affiliation{Wigner Research Centre for Physics, Institute for Solid State
Physics and Optics, H-1525 Budapest, P.O.Box 49, Hungary}
\affiliation{Institute of Theoretical Physics, Szeged University, H-6720 Szeged,
Hungary}
\affiliation{Center for Complex Networks Research and Department of Physics,
Northeastern University, 177 Huntington Avenue,
Boston, MA 02115, USA}
\author{Gerg\H o Ro\'osz}
\email{roosz.gergo@wigner.mta.hu}
\affiliation{Wigner Research Centre for Physics, Institute for Solid State Physics and Optics, H-1525 Budapest, P.O.Box 49, Hungary}
\affiliation{Institute of Theoretical Physics, Szeged University, H-6720 Szeged, Hungary}
\author{Ferenc Igl\'oi}
\email{igloi.ferenc@wigner.mta.hu}
\affiliation{Wigner Research Centre for Physics, Institute for Solid State
Physics and Optics, H-1525 Budapest, P.O.Box 49, Hungary}
\affiliation{Institute of Theoretical Physics, Szeged University, H-6720 Szeged,
Hungary}
\date{\today}

\begin{abstract}
The entanglement entropy in clean, as well as in random quantum spin chains has a logarithmic size-dependence at the critical point.
Here, we study the entanglement of composite systems that consist of a clean and a random part, both being critical. In the composite, antiferromagnetic XX-chain with a sharp interface, the entropy is found to grow in a double-logarithmic fashion ${\cal S}\sim \ln\ln(L)$, where $L$ is the length of the chain. We have also considered an extended defect at the interface, where the disorder penetrates into the homogeneous region in such a way that the strength of disorder decays with the distance $l$ from the contact point as $\sim l^{-\kappa}$. For $\kappa<1/2$, the entropy scales as ${\cal S}(\kappa) \simeq (1-2\kappa){\cal S}(\kappa=0)$, while for $\kappa \ge 1/2$, when the extended interface defect is an irrelevant perturbation, we recover the double-logarithmic scaling. These results are explained through strong-disorder RG arguments.
\end{abstract}

\maketitle

\section{Introduction}
\label{sec:intr}
Recently, we are witnessing an increasing theoretical interest in the entanglement properties of many-body quantum systems, for reviews see\cite{amico,entanglement_review,area,laflorencie}.
In a quantum system, being in a pure state $|\Psi\rangle$ and having a density matrix $\rho=| \Psi\rangle\langle\Psi |$, the entanglement between its confined part ${\cal A}$ and its complement ${\cal B}$ can be quantified by the entanglement entropy\cite{bennett}, which is just the von Neumann entropy of either subsystem:
\be
{\cal S}_{\cal A}=-{\rm Tr}(\rho_{\cal A} \ln \rho_{\cal A}) = 
           -{\rm Tr}(\rho_{\cal B} \ln \rho_{\cal B}) =
{\cal S}_{\cal B}\;.
\label{S}
\ee
Here, $\rho_{\cal A}={\rm Tr}_{\cal B}\rho$ and  $\rho_{\cal B}={\rm Tr}_{\cal A}\rho$ are reduced density matrices.
Other entanglement measures, such as the R\'enyi entropies\cite{renyi} can be defined similarly.

Most of the studies are restricted to one-dimensional systems, in which the subsystem ${\cal A}$ (${\cal B}$) is represented by the sites $i=1,2,\dots,\ell$ ($i=\ell+1,\ell+2,\dots,L$). In a gapped state the entanglement entropy approaches a finite limiting value, as $L \to \infty$ and $\ell \to \infty$, which is the special case of the so-called area law. In a critical state, however, in which there is a quasi-long-range order with algebraically decaying correlations, the entanglement entropy is diverging. In conformally invariant systems, the divergence is logarithmic\cite{holzhey,Calabrese_Cardy04,vidal,peschel03,jin_korepin,peschel04,IJ07}:
\be
S_{\cal A}(\ell) \simeq \frac{c}{6}b \ln \ell + const\;,
\label{c}
\ee
where $c$ is the central charge of the conformal algebra and $b$ is the number of contact points between
$\mathcal{A}$ and $\mathcal{B}$. We note that there are examples of quantum spin chains, which are not conformally invariant, where the entanglement entropy grows faster than logarithmically\cite{growth}.

Interestingly, the logarithmic divergence of the entanglement entropy holds to be valid for random quantum spin chains, the properties of which are controlled by infinite-disorder fixed points\cite{refael,Santachiara,Bonesteel,s=1,Laflo05,IgloiLin08,dyn06}. Examples are the antiferromagnetic (AF) random Heisenberg and XX-chains, as well as the random transverse-field Ising chain (TFIC) at the critical point. The scaling behavior of the average entanglement entropy\cite{refael}, as well as its distribution function\cite{huse} have been calculated analytically by the strong-disorder renormalization group (SDRG) method\cite{im}.

There are only a few results about the entanglement properties of composite systems, in which the two parts ${\cal A}$ and ${\cal B}$ have different properties. Here we mention a study of composite free-fermion chains\cite{peschel}, in which the two halves are either metallic, having a vanishing gap, or insulating. In the metal-metal system the logarithmic divergence in Eq.(\ref{c}) stays valid, with an effective central charge, which depends only on the value of the connecting coupling. On the contrary, for metal-insulator or insulator-insulator systems the area law remains valid.

In the present paper, we consider composite quantum spin chains, in which the strength of quenched disorder is different in different parts of the system. In particular, we consider the AF XX-chain and calculate the entanglement entropy between random and non-random parts. This problem can be equivalently formulated in terms of critical transverse-field Ising chains, too\cite{IJ07}. The interface region separating the two parts of the system is either sharp, or the disorder penetrates into the homogeneous part in such a way, that the strength of disorder decreases with the distance from the contact point as $l^{-\kappa}$\cite{note}. For $\kappa \to \infty$, we formally recover the sharp interface, while $\kappa=0$ corresponds to the usual, homogeneously disordered system. With increasing $\kappa$, the interface becomes sharper and sharper. We use scaling results, strong-disorder RG arguments and exact results to show that, for $\kappa>1/2$, the extended interface defect is an irrelevant perturbation and the interface asymptotically behaves as a localised one. We have studied the entanglement properties of this composite systems by two methods: numerically through free-fermion techniques\cite{vidal,peschel03,jin_korepin,peschel04,IJ07} and, for the extended defect with $\kappa \le 1/2$, by a numerical implementation of the SDRG approach. By the latter method, which is expected to provide asymptotically exact results, generally much longer chains can be studied.

The rest of the paper is organised as follows. The AF XX-chain and the related critical transverse-field Ising chain is introduced in Sec.\ref{sec:model}. In Sec.\ref{sec:inhom}, we consider an extended interface defect and first we study its local critical behaviour through scaling arguments, by calculating the local order parameter and through an SDRG approach. Then the entanglement entropy across an extended defect is calculated in the composite system. In Sec.\ref{sec:disorder}, the entanglement entropy of the composite system with a sharp interface is considered. Finally, our results are discussed in Sec.\ref{sec:discussion}. Some details of the calculations are presented in the Appendices. 

%%%%%%%%%%%%%%%%%%%%%%%%%%%%%%%%%%%%%%%%%%%%%%%%%%%%%%%%%%%%%%%%%%%%%%%%%%%%
\section{Models and methods}
\label{sec:model}
We consider a composite AF XX-chain defined by the Hamiltonian:
\be
{\cal H}_{XX} = {\cal H}_{XX}^{({\cal A})} + {\cal H}_{XX}^{({\cal B})} + {\cal V}_0+{\cal V}_{2L}\;.
\label{eq:H_XX}
\ee
Here, ${\cal H}_{XX}^{({\cal A})}$ and ${\cal H}_{XX}^{({\cal B})}$ represent two open chains of length $L$:
\be
{\cal H}_{XX}^{({\cal A})} =
\sum_{1 \le i \le L-1} J_i(S_i^x S_{i+1}^x+S_i^y S_{i+1}^y)\;,
\label{eq:H_XX_A}
\ee
in terms of spin-$1/2$ operators $S_i^{x,y}$ at site $i$. In ${\cal H}_{XX}^{({\cal B})}$, the summation is performed for $-L \le i \le -2$. The junction terms are ${\cal V}_0=J_0S_{-1}^x S_{1}^x$ and ${\cal V}_L=J_L S_{-L}^x S_{L}^x$. 

Subsystem ${\cal A}$ is homogeneously disordered, the couplings are taken randomly from a uniform distribution in the interval $(1-\Delta,1+\Delta)$, $0\le\Delta \le 1$. Subsystem ${\cal B}$ is non-random at least in its bulk, i.e. asymptotically far from the contact points, with unit couplings. The two parts are separated by either a sharp interface at $i=0$, or ${\cal B}$ contains an extended defect, where a coupling at a distance $l$ from the closest contact point ($l \le L/2$) is taken from a uniform distribution in the interval $(1-\Delta_l,1+\Delta_l)$. Here, $\Delta_l$ tends to zero with increasing $l$ asymptotically as
$\Delta_l \sim l^{-\kappa}$, $\kappa \ge 0$. In the limiting case, $\kappa=0$, the subsystem is homogeneously disordered and with increasing $\kappa$, it interpolates between the homogeneously random and the non-random system. We are interested in the entanglement entropy between the two parts of the system.

The above problem can be equivalently formulated in terms of transverse-field Ising chains (TFIC-s) by using a well-known mapping\cite{peschel_schotte,fisherxx,ijr00}. First, let us define two sets of Pauli operators, $\sigma_i^{x,z}$ and $\tau_i^{x,z}$, $-L/2 \le i \le L/2$
through the spin-1/2 operators $S_j^{x,y}$, by:
\beqn
\sigma_i^x=\prod_{j=-L}^{2i-1} (2S_j^x),\quad \sigma_i^z=4 S_{2i-1}^y S_{2i}^y \cr
\tau_i^x=\prod_{j=-L}^{2i-1} (2S_j^y),\quad \tau_i^z=4 S_{2i-1}^x S_{2i}^x\;.
\label{mapping}
\eeqn
In terms of these Pauli operators, the Hamiltonian of the XX chain in
Eq.(\ref{eq:H_XX}) can be written as the sum of two decoupled TFIC-s. We have for the Hamiltonian of chain ${\cal A}$ in Eq.(\ref{eq:H_XX_A}):
\beqn
{\cal H}_{XX}^{({\cal A})}&=&\frac{1}{2}[{\cal H}_I^{({\cal A})}(\sigma)+{\cal H}_I^{({\cal A})}(\tau)]\cr
{\cal H}_I^{({\cal A})}(\sigma)&=&-\frac{1}{2}\sum_{i=1}^{L/2} J_{2i}\sigma_i^x \sigma_{i+1}^x-
\frac{1}{2}\sum_{i=1}^{L/2} J_{2i-1} \sigma_i^z\;,
\label{mapping_H}
\eeqn
and ${\cal H}_I^{({\cal A})}(\tau)$ is in the same form as ${\cal H}_I^{({\cal A})}(\sigma)$. As shown in Ref.\cite{IJ07}, the entanglement entropy of the two systems are related as
\be
{\cal S}^{(XX)}={\cal S}^{(\sigma)}+{\cal S}^{(\tau)}\;,
\label{S_rel}
\ee
where ${\cal S}^{(\sigma)}={\cal S}^{(\tau)}$.

The entanglement entropy of the XX chain can be calculated by free-fermion methods, which necessitates the solution of the eigenvalue problem of $L \times L$ matrices. This can be achieved numerically by ${\cal O}(L^3)$ operations. In a random AF XX-chain, the asymptotic $L$-dependence of the entanglement entropy can be calculated by the numerical implementation of the SDRG method. This numerical procedure is faster, requiring only ${\cal O}(L \ln L)$ operations for a random sample. Finally, for a composite (clean and random) XX chain, we can use an approximate mixed procedure. As described in Sec.\ref{sec:disorder}, first the random part is renormalised through the local use of the SDRG rules, which results in a chain of ${\cal O}(\ln L)$ spins with random couplings. The computation of the entanglement entropy of this part with the clean one of length $L$ through the free-fermion method can be performed by ${\cal O}(L^2)$ operations, which is described in Appendix \ref{appendix_B}.

%%%%%%%%%%%%%%%%%%%%%%%%%%%%%%%%%%%%%%%%%%%%%%%%%%%%%%%%%%%%%%%%%%%%%%%%%%%
\section{Extended defect at the interface}
\label{sec:inhom}

In this section, we consider the composite system in which the random and non-random parts are separated by an extended defect. First, we study the effect of the extended defect on the local critical behaviour of the non-random system. This question is analysed through a relevance-irrelevance criterion, by direct calculation of the surface order parameter and through a strong-disorder RG approach. In the second part of the section, the entanglement entropy of this composite system is studied numerically.

\subsection{Relevance-irrelevance criterion}
\label{sec:harris}

Let us study the relevance or irrelevance of the extended defect on the local critical behaviour of the non-random chain, i.e. when ${\cal B}$ is separated from ${\cal A}$. To do so, we generalise the Harris criterion\cite{harris} for homogeneously disordered chains and consider a correlated domain at the surface of ${\cal B}$ with an extent $\xi$.
The energy gain per site due to disorder fluctuations in this domain is given by 
\be
\epsilon_{dis} \sim \frac{1}{\xi}\left[\sum_{l=1}^{\xi} \Delta_l^2 \right]^{1/2} \sim \xi^{-1/2-\kappa}\;,
\ee
which is to be compared with the energy gap $\epsilon_q$ related to quantum fluctuations in the clean, off-critical system. Off-criticality can be achieved in the XX chain by introducing a dimerization by couplings $J_{2i}=1+\delta$ and $J_{2i-1}=1-\delta$, which results in $\epsilon_q\sim\delta$. (This reasoning follows also from the mapping to the TFIC, see Eq.(\ref{mapping_H}), where $\delta$ is the control parameter measuring the distance from the critical point.) The size of the correlated domain is just the correlation length in the clean system: $\xi \sim \delta^{-\nu}$, where $\nu=1$ is the correlation-length critical exponent of the XX chain. The ratio of the two types of energy corrections is
\be
\frac{\epsilon_{dis}}{\epsilon_{q}} \sim \delta^{\phi},\quad \phi=\nu(1/2+\kappa)-1\;,
\ee
which, at the critical point ($\delta \to 0$), goes to zero for $\phi>0$. In this case, the extended defect is an irrelevant perturbation. This means that, for $\kappa>1/2$, the surface critical behaviour is the same as in the clean system. If, however, $\phi<0$, the perturbation is relevant and consequently, for $\kappa<1/2$, the local critical behaviour of the system at the surface is different from that of the clean system.

\subsection{Surface order parameter}

Let us consider now the order parameter at the boundary, which is defined as $m_1^x=\langle 0 | S_l^x | 0 \rangle$ and can be obtained in a closed form even for finite chains if we use
fixed spin boundary condition, $S_L^x=\pm 1/2$, which amounts to have
$J_{L-1}=0$. As shown in Ref.\cite{ijr00}
\be
m_1^x=\displaystyle
{1 \over 2} \left[1+\sum_{l=1}^{L/2-1} \prod_{j=1}^l
\left( J_{2j-1} \over J_{2j} \right)^2\right]^{-1/2}\;.
\label{peschel}
\ee
The typical behaviour of $m_1^x(L)$ is different for $\kappa<1/2$ and for $\kappa>1/2$. In the former case, the typical value of $m_1^x(L)$ is dominated by the largest product, say at $l=\ell$: $\prod_{j=1}^{\ell} \left( J_{2j-1} \over J_{2j} \right)^2 \sim \exp\left[2\sum_{j=1}^{l}\theta_j\right]$, with $\theta_j=\ln J_{2j-1} - \ln J_{2j-1}$. Here $\theta_j$ is a random number of zero mean and (for large-$j$) of variance $\sim j^{-\kappa}$, thus the dominant product scales as $\sim \exp\left[c \ell^{(1-2\kappa)/2} \right]$. In this case, the \textit{typical} value of the surface order parameter is given by $(\ell \sim L)$:
\be
m_{1,{\rm typ}}^x(L) \sim \exp\left[-c L^{(1-2\kappa)/2} \right], \quad \kappa<1/2\;.
\label{m_s_typ}
\ee
The average value of the order parameter is determined by rare realizations, in which all products in Eq.(\ref{peschel}) are negligible, and thus $m_{1,{\rm rare}}^x(L)={\cal O}(1)$.  As in the case of homogeneous disorder, the rare realizations can be identified as random walks (with decreasing step length) which have a surviving character\cite{ijr00}.

On the contrary, for $\kappa>1/2$ each product in the sum of Eq.(\ref{peschel}) is of ${\cal O}(1)$ and then 
\be
m_{1,{\rm typ}}^x(L) \sim L^{-1/2}, \quad \kappa>1/2\;,
\ee
which is in the same form as in the pure system. Thus the perturbation is irrelevant in this case in agreement with the Harris criterion.

Considering the marginal situation, $\kappa=1/2$, the typical value of the surface order parameter behaves for large-$L$ as:
\be
m_{1,{\rm typ}}^x(L) \sim L^{-1/2}\exp\left(-|a|\sqrt{\ln L}\right), \quad \kappa=1/2\;.
\ee
This means that there is a logarithmic correction to the clean system critical behaviour, thus the perturbation is marginally irrelevant.

\subsection{Strong-disorder RG approach}
\label{sec:SDRG}

The critical properties of the homogeneously disordered XX chain can be studied by the strong-disorder RG method\cite{im,fisherxx,mdh}. Here, we recapitulate the basic results and suggest their generalisation for inhomogeneous disorder. 
By the SDRG method, an approximate ground state of the random XX chain can be constructed, which becomes asymptotically exact on large scales (low energy scales). The elementary step of the procedure is the projection of the local state of a block of adjacent spins connected by the maximal coupling $\Omega=\max\{J_i\}$ onto a singlet state. This is justified if the neighbouring two couplings are much smaller than $\Omega$, $J_1,J_3\ll J_2=\Omega$, 
and in this case, second-order perturbation theory gives a weak effective antiferromagnetic coupling $\tilde J=J_1J_3/\Omega$ between spins next to the singlet. Applying this reduction step iteratively for the actually maximal coupling, the energy scale $\Omega$ gradually decreases, while the above condition is fulfilled more and more accurately and, starting with a finite, even number of spins, the procedure ends up with a product of pairs of spins in a singlet state, which is called random singlet state.      
In such a state, the entanglement entropy is given by the number of spin pairs with one constituent in subsystem $\mathcal{A}$ and the other one in subsystem $\mathcal{B}$, multiplied by $\ln 2$.  
For any initial distribution of the reduced couplings  $\beta=\ln(\Omega/J)$ with a variance $O(1)$, the attractor for the distribution of $\beta$ at low energy scales is known to be of the form\cite{im,fisherxx}:
\be  
P_{\Gamma}(\beta)=\frac{1}{\Gamma+\Gamma_0}e^{-\beta/(\Gamma+\Gamma_0)}.
\label{P}
\ee
Here $\Gamma=\ln(\Omega_0/\Omega)$, where $\Omega_0$ is the initial value of $\Omega$, is a logarithmic energy scale, and $\Gamma_0$ is an initial strength of the disorder.
The density of non-decimated, i.e. active spins, $c(\Gamma)$, obeys the differential equation:
\be 
-\frac{dc}{c}=2\frac{1}{\Gamma+\Gamma_0}d\Gamma,
\label{diff}
\ee
with the solution:
\be 
c(\Gamma)=\left(\frac{\Gamma}{\Gamma_0}+1\right)^{-2}.  
\label{xi}
\ee
From this, one obtains the renormalization length scale, i.e. the average distance between active spins by $\xi = 1/c$. The mean number of singlet pairs connecting the two subsystems generated when the logarithmic energy scale changes from $\Gamma$ to $\Gamma+d\Gamma$ is given by\cite{refael} 
\be
dN(\Gamma)=\frac{1}{3}\frac{1}{\Gamma+\Gamma_0}d\Gamma.
\label{dN}
\ee
Integrating up to scale $\Gamma$ leads to
\be 
N(\Gamma)=
%\int_0^{\Gamma}p_0(\Gamma)d\Gamma=
\frac{1}{3}\ln\Gamma+const.
%=\frac{1}{6}\ln\xi. 
\label{NGamma}
\ee
Using Eq. (\ref{xi}) with the setting $\xi=L$, this formula gives for the asymptotic size dependence of average entanglement entropy in a homogeneously disordered chain ${\cal S}_L=\frac{c_{\rm eff}}{6}\ln L + const.$, with $c_{\rm eff}=\ln 2$.

Now let us return to an infinitely large composite system with identically distributed disorder in part $\mathcal{A}$ and an extended interface disorder in part $\mathcal{B}$. In $\mathcal{B}$ the fixed-point distribution of the  reduced couplings  $\beta_l=\ln(\Omega/J_l)$ is given in a similar form as in Eq.(\ref{P}), however the initial disorder parameter $\Gamma_0$ is replaced by some site-dependent one, $\Gamma_l$:
\be  
P_{\Gamma}(\beta)=\frac{1}{\Gamma+\Gamma_l}e^{-\beta/(\Gamma+\Gamma_l)}.
\label{Pgamma_l}
\ee
The value of $\Gamma_l$ can be obtained by calculating the density of active sites, $c_l$, which according to Eq.(\ref{xi}) for large-$\Gamma$ scales as: $c_l\approx c (\Gamma_l/\Gamma_0)^2$.

To obtain an estimate for $c_l$ we introduce a modified SDRG procedure, in which, at the initial problem, we divide the system into $l$-dependent blocks of spins having a length $b_l \sim l^{2\kappa}$ and any two blocks are separated by an extra spin. In the first step of the renormalization each block is eliminated independently. To obtain the renormalized value of the coupling between the remaining spins, ${\tilde J}_l$, we calculate for the given block the ground-state energy with fixed spin boundary conditions: $E_0^{\uparrow \uparrow}$ and $E_0^{\uparrow \downarrow}$. In the perturbation calculation, their difference is given by ${\tilde J}_l/2$, which is ${\cal O}(1)$ independently of $l$. This statement follows from the approximate formula for the excitation energy in Ref.\cite{ijr00}, or can be derived by perturbation calculation. 
In this way, we arrive at a disordered chain with site independent disorder of variance ${\cal O}(1)$, which is then renormalized further with the usual SDRG method. The number of blocks, $B$, in a chain of length $L$ scales as $B \sim L^{1-2\kappa}$, which means that an infinite-disorder fixed point (IDFP) can exist only for $\kappa<1/2$. Interestingly, the stability limits of the IDFP and that of the pure system's fixed point coincide for the present model.

Returning to the density of active sites, we obtain
\be
c_l \sim \frac{\Gamma^{-2}}{b_l} \sim \Gamma^{-2}l^{-2\kappa},\quad \kappa<1/2\;,
\ee
thus $\Gamma_l \sim l^{-\kappa}$.
We can see in Eq. (\ref{Pgamma_l}) that the site-dependence of couplings enters through $\Gamma_l$, but for large scales, $\Gamma\gg \Gamma_0>\Gamma_l$, this position dependence vanishes. This means that disregarding the length of bonds and focussing only on the couplings, the renormalized system becomes homogeneously distributed on asymptotically large scales. 
We can therefore conclude that the asymptotic form in Eq. (\ref{NGamma}), which is written in terms of $\Gamma$ is valid also for the composite system.
The difference compared to the case of identically distributed disorder ($\kappa=0$) appears if we translate this expression to $L$-dependence.  
Considering a finite system of size $L$, the formation of singlet pairs connecting the two subsystems stops at the scale $\Gamma_L$ at which the mean number of active spins remaining in part $\mathcal{B}$ is reduced to ${\cal O}(1)$:
\be 
n_T(\Gamma)\sim \int_{}^{L}c_ldl \sim \Gamma^{-2}L^{1-2\kappa} \sim {\cal O}(1),\quad \kappa<1/2\;,
\ee
which yields
\be
\Gamma_L\sim L^{(1-2\kappa)/2}.
\ee
Substituting this into Eq. (\ref{NGamma}), we obtain for the $L$-dependence of the average entanglement entropy 
\be 
{\cal S}_L \simeq \frac{c_{\rm eff}(\kappa)}{6} \ln L, \qquad (\kappa<1/2)\;,
\label{S_kappa}
\ee 
with an effective central charge: $c_{\rm eff}(\kappa)=\ln 2(1-2\kappa)$, which is reduced and varies continuously with $\kappa$.
In the limiting case $\kappa=1/2$ we have 
$n_T(\Gamma)\sim \Gamma^{-2}\ln L\sim O(1)$, which results in a double-logarithmic dependence of the entropy: 
$S_L \sim \ln\ln L$.

%%%%%%%%%%%%%%%%%%%%%%%%%%%%%%%%%%%%%%%%%%%%%%%%%%%%%%%%%%%%%%%%%%%%%%%%%%%
\subsection{Entanglement entropy: numerical results}
\label{sec:numerical}

We have performed numerical SDRG calculations and exact diagonalization by the free-fermion technique using the following distribution. 
In part ${\cal A}$, the couplings are i.i.d. and drawn from a uniform distribution $[0,1]$. In part ${\cal B}$, $J_i$ is drawn from a uniform distribution with a support $[(1-\Delta_i)/2,(1+\Delta_i)/2]$, where $\Delta_i=(i+1)^{-\kappa}$. 
We have calculated the average entanglement entropy between the two parts by both methods. 
The number of samples was $10^7$ in the SDRG calculations and $10^6$ in the diagonalization method. Periodic chains were used with total length $2L$.

%%%%%%%%%%%%%%%%%%%%%%%%%%%%%%%%%%%%%%%%%%%%%%%%%%%%%%%%%%%%%%%%%%%%%%%%%
\begin{figure}[h!]
\includegraphics[width=9cm]{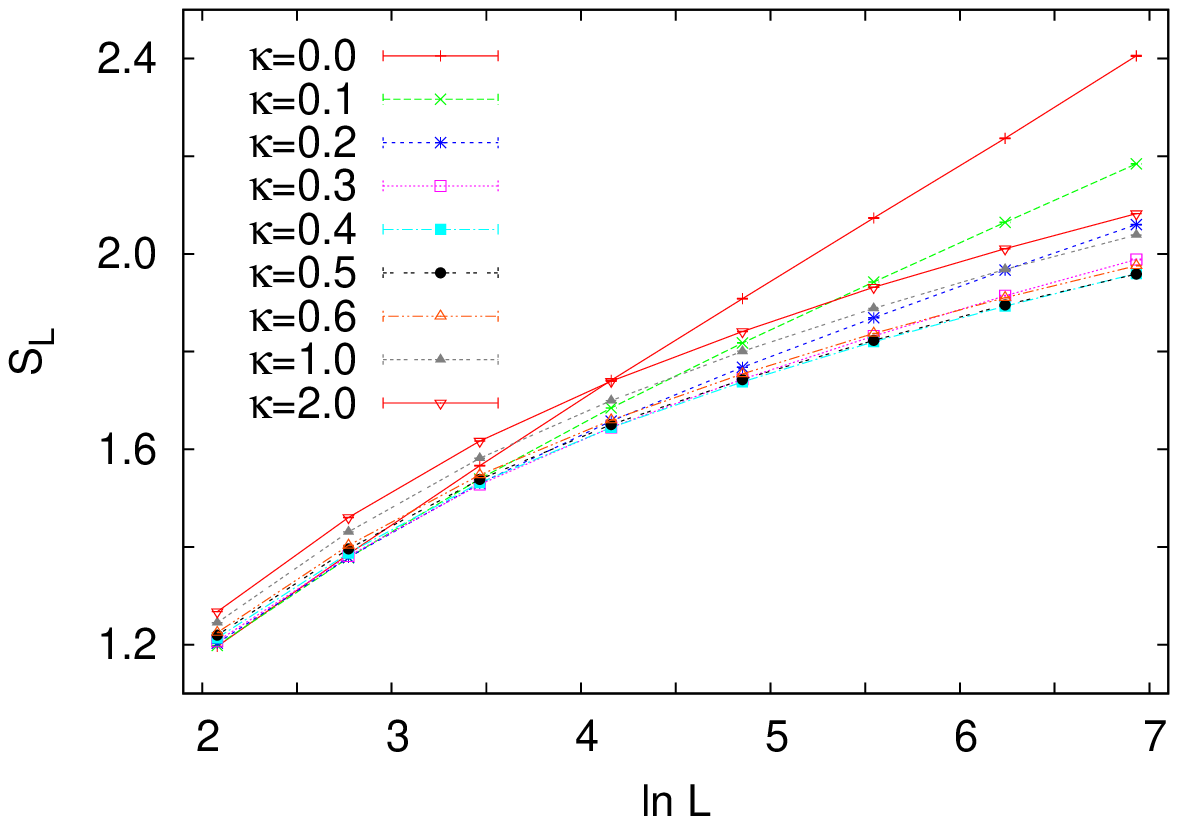} 
\includegraphics[width=9cm]{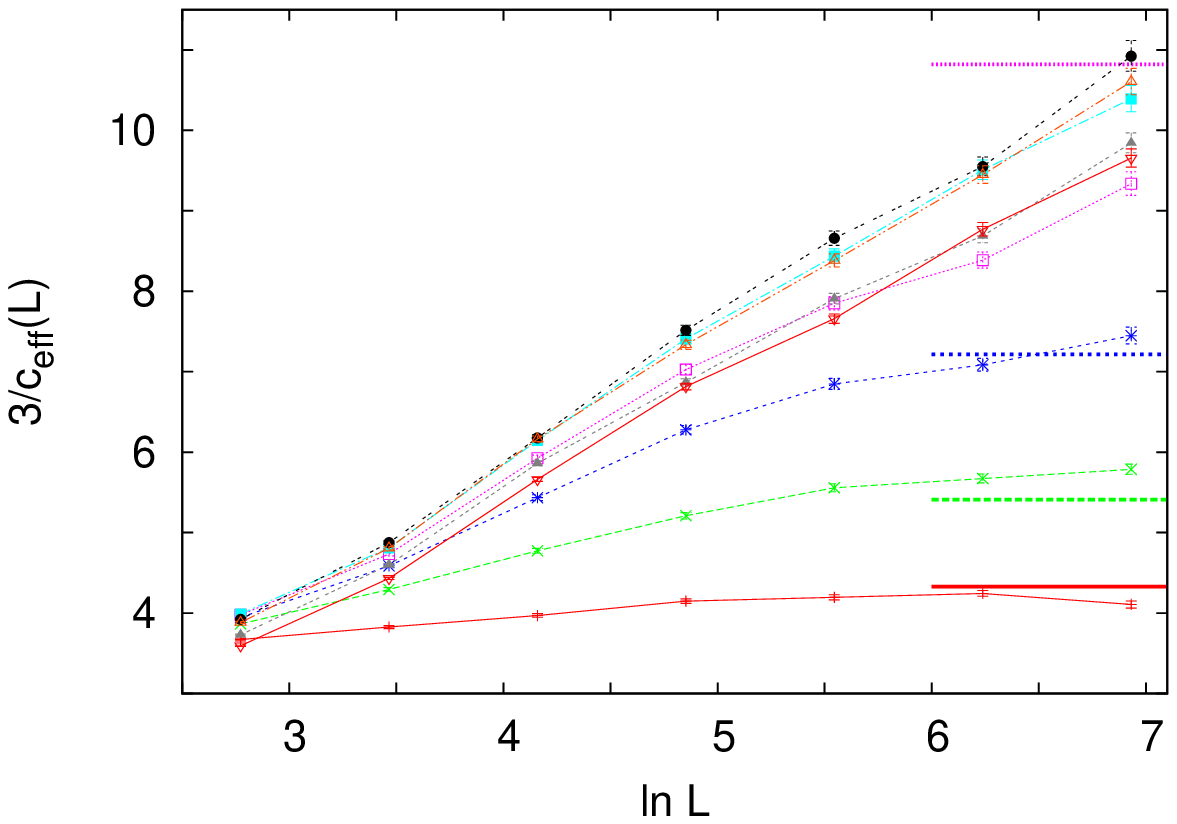} 
\caption{\label{mixper} (Color online) Top. Size-dependence of the average entanglement entropy in the composite AF XX-chain with an extended interface defect obtained by free-fermion methods for different values of $\kappa$. 
Bottom: Inverse of the effective central charge obtained through two-point fits for the same values of $\kappa$ as above. The horizontal lines indicate the conjectured value by the SDRG approach.  
}
\end{figure}
%%%%%%%%%%%%%%%%%%%%%%%%%%%%%%%%%%%%%%%%%%%%%%%%%%%%%%%%%%%%%%%%%%%%%%%%%%%%%%%%%%%%%%%%%%%%%%%%%%%%%%%%%%%%%%%%%%%%%%%%%%%%%%%%%%%%%%%%%%%%%%%%%%%

\begin{figure}[h!]
\includegraphics[width=9cm]{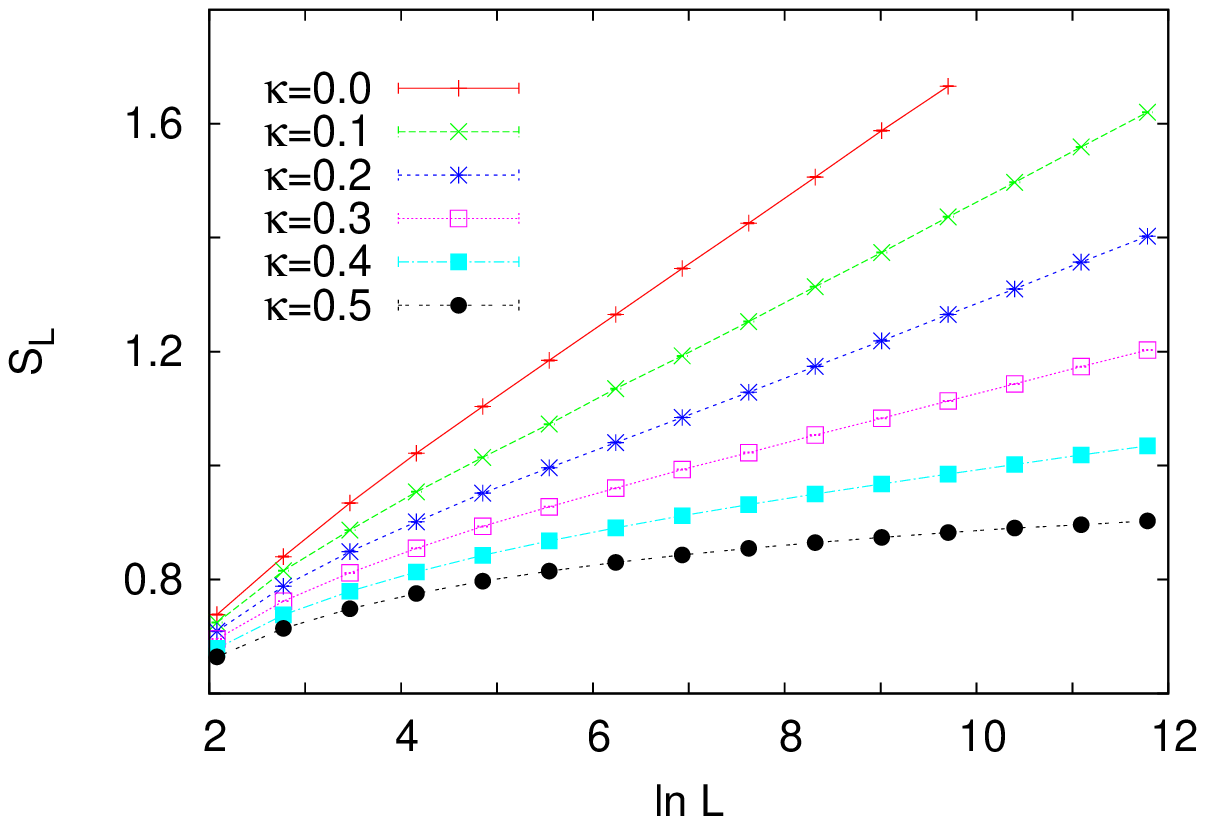} 
\includegraphics[width=9cm]{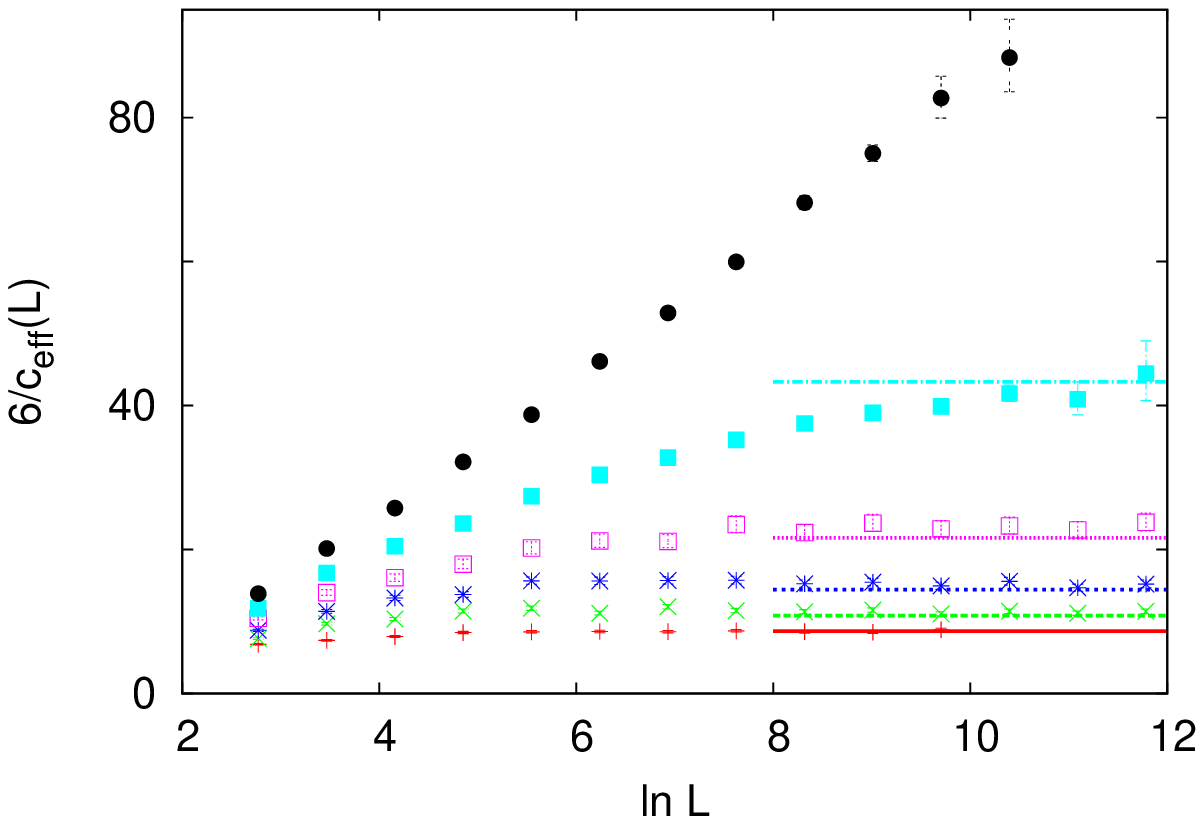} 
\caption{\label{mixrg} (Color online) The same as in Fig.\ref{mixper} calculated by the numerical application of the SDRG method for $\kappa \le 1/2$.  
}
\end{figure}

The entropy obtained by the free-fermion method is shown in Fig.\ref{mixper} as a function of $\ln L$ for different values of $\kappa$.
The curves seem to approach a linear asymptotic behaviour for $\kappa < 0.5$ and the slope is given by $c_{\rm eff}/3$. For the effective
central charge, we have the conjectured value by the SDRG method: $c_{\rm eff}=\ln 2 (1-2\kappa)$, see in Sec.\ref{sec:SDRG}. Through two-point fits, we have calculated size-dependent estimates for $c_{\rm eff}(L)$, which are shown in Fig.\ref{mixper}. For not too large values of $\kappa \le 0.2$ the numerical results are in agreement with the conjectured formulae. For larger values, $\kappa=0.3$ and $0.4$ the available system sizes are too small to obtain asymptotic results. In theses cases the conjectured relations are verified by SDRG studies, see in Fig.\ref{mixrg}. At the borderline case, $\kappa=0.5$, the effective central charge
vanishes as $c_{\rm eff}(L) \sim 1/\ln L$, which agrees with the double-logarithmic size-dependence of the entropy. For larger values of $\kappa>1/2$, $c_{\rm eff}(L)$ shows the same type of logarithmic dependence, which means that the double-logarithmic size-dependence of the entanglement entropy remains valid in this case, too. The prefactor of the double-logarithm is approximately constant for $\kappa > 1/2$, and estimated to be $\gamma \approx 0.35(2)$ for one contact point.

%%%%%%%%%%%%%%%%%%%%%%%%%%%%%%%%%%%%%%%%%%%%%%%%%%%%%%%%%%%%%%%%%%%%%%%%%
\begin{figure}[h!]
\includegraphics[width=9cm]{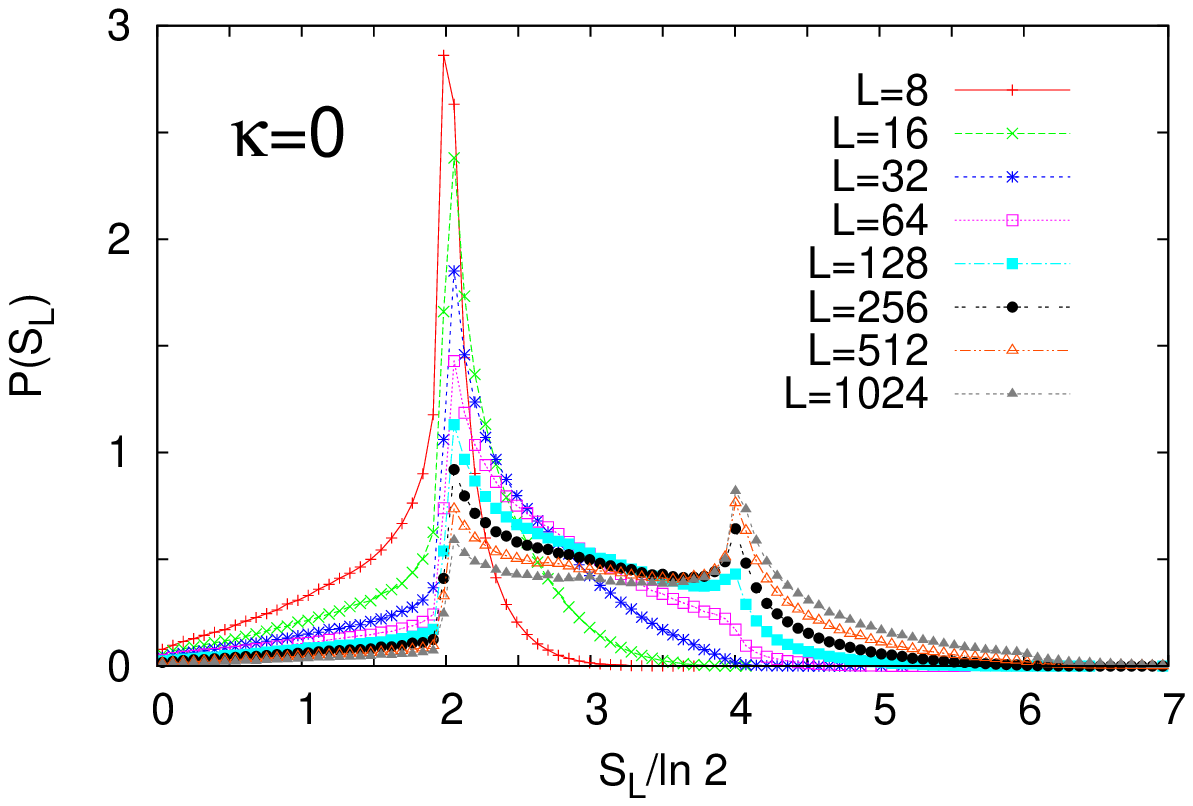} 
\includegraphics[width=9cm]{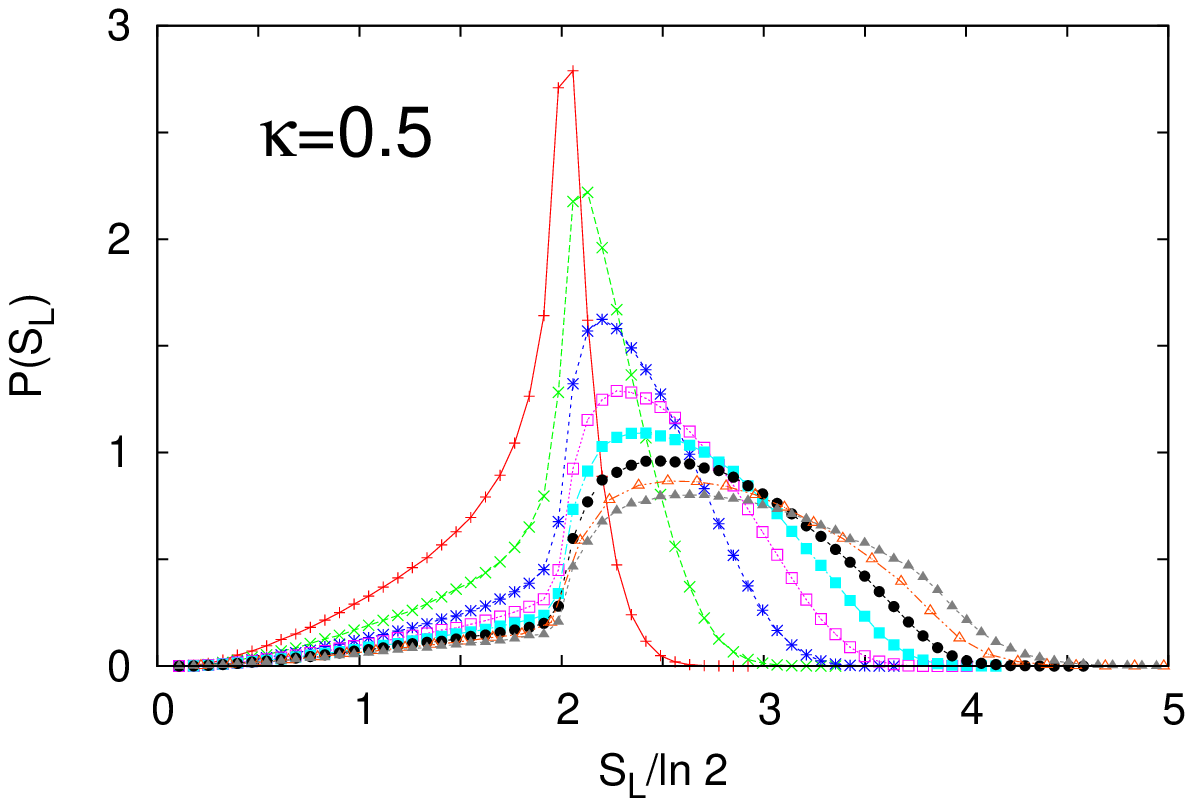}
\includegraphics[width=9cm]{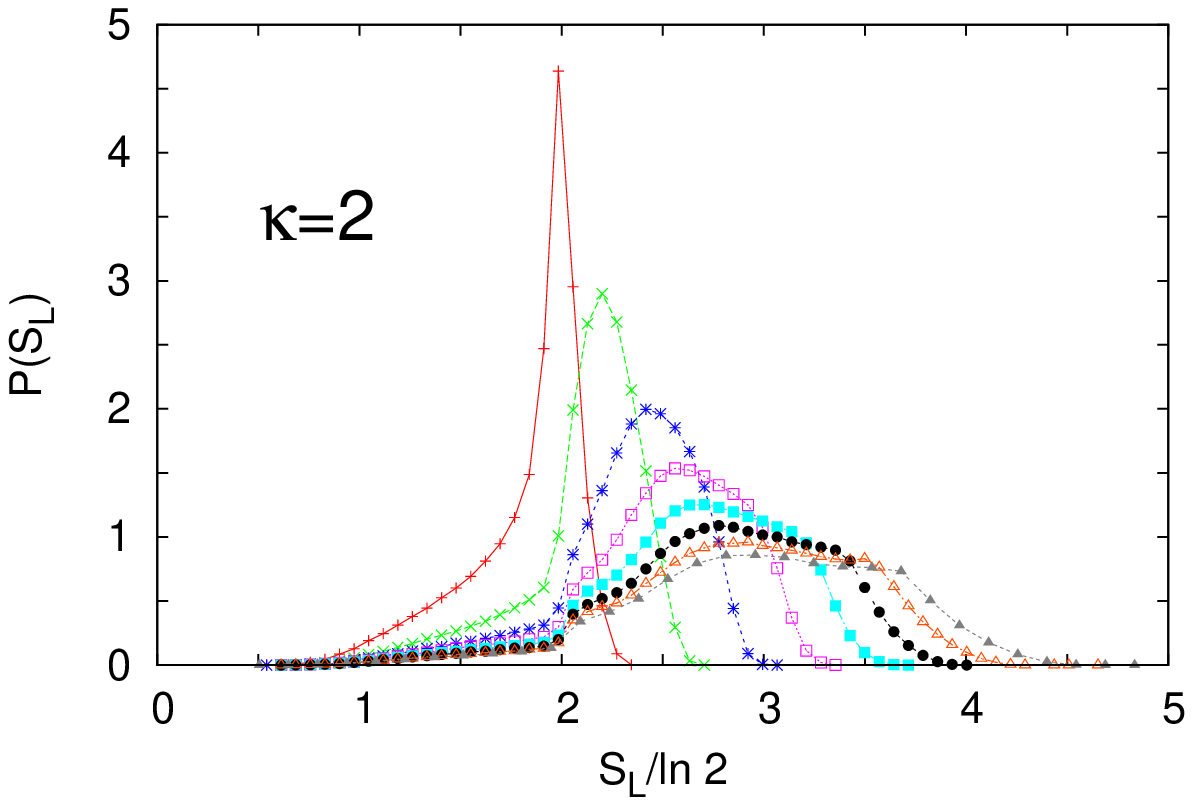}
\caption{\label{sdist} (Color online) Distribution of the entanglement entropy in the composite AF XX-chain with an extended interface defect obtained by the free-fermion method. Top: two homogeneously disordered parts ($\kappa=0$); middle: $\kappa=0.5$; 
bottom: $\kappa=2.0$.  
}
\end{figure}
%%%%%%%%%%%%%%%%%%%%%%%%%%%%%%%%%%%%%%%%%%%%%%%%%%%%%%%%%%%%%%%%%%%%%%%%%%%%%%%%%%%%%%%%%%%%%%%%%%%%%%%%%%%%%%%%%%%%%%%%%%%%%%%%%%%%%%%%%%%%%%%%%%%
In Fig.\ref{sdist}, we show the distribution of the entanglement entropy obtained by the free-fermion method for different values of $\kappa$. In the case of two homogeneously disordered parts ($\kappa=0$), there are peaks at ${S(L)}{\ln 2}=2$ and $4$ (as well as at $6$), which correspond to one and two singlets across the two junctions. For the double-logarithmic average entropy cases $\kappa=0.5$ and $\kappa=2.0$, the distributions are more flat.

%%%%%%%%%%%%%%%%%%%%%%%%%%%%%%%%%%%%%%%%%%%%%%%%%%%%%%%%%%%%%%%%%%%%%%%%%
%%%%%%%%%%%%%%%%%%%%%%%%%%%%%%%%%%%%%%%%%%%%%%%%%%%%%%%%%%%%%%%%%%%%%%%%%%%%%%%%%%%%%%%%%%%%%%%%%%%%%%%%%%%%%%%%%%%%%%%%%%%%%%%%%%%%%%%%%%%%%%%%%%%
The average entanglement entropy and the size-dependent effective central charges are calculated by the traditional use of the SDRG method, too, the results are shown in Fig.\ref{mixrg}. The curves are very similar to that in Fig.\ref{mixper}, however larger systems have been treated with better statistics. As a result, the calculated size-dependent effective central charges are in good agreement with the analytical SDRG conjecture, even for $\kappa=0.3$ and $0.4$.

%%%%%%%%%%%%%%%%%%%%%%%%%%%%%%%%%%%%%%%%%%%%%%%%%%%%%%%%%%%%%%%%%%%%%%%%%%%
\section{Sharp interface}
\label{sec:disorder} 
\begin{figure}[h]
\includegraphics[width=9cm]{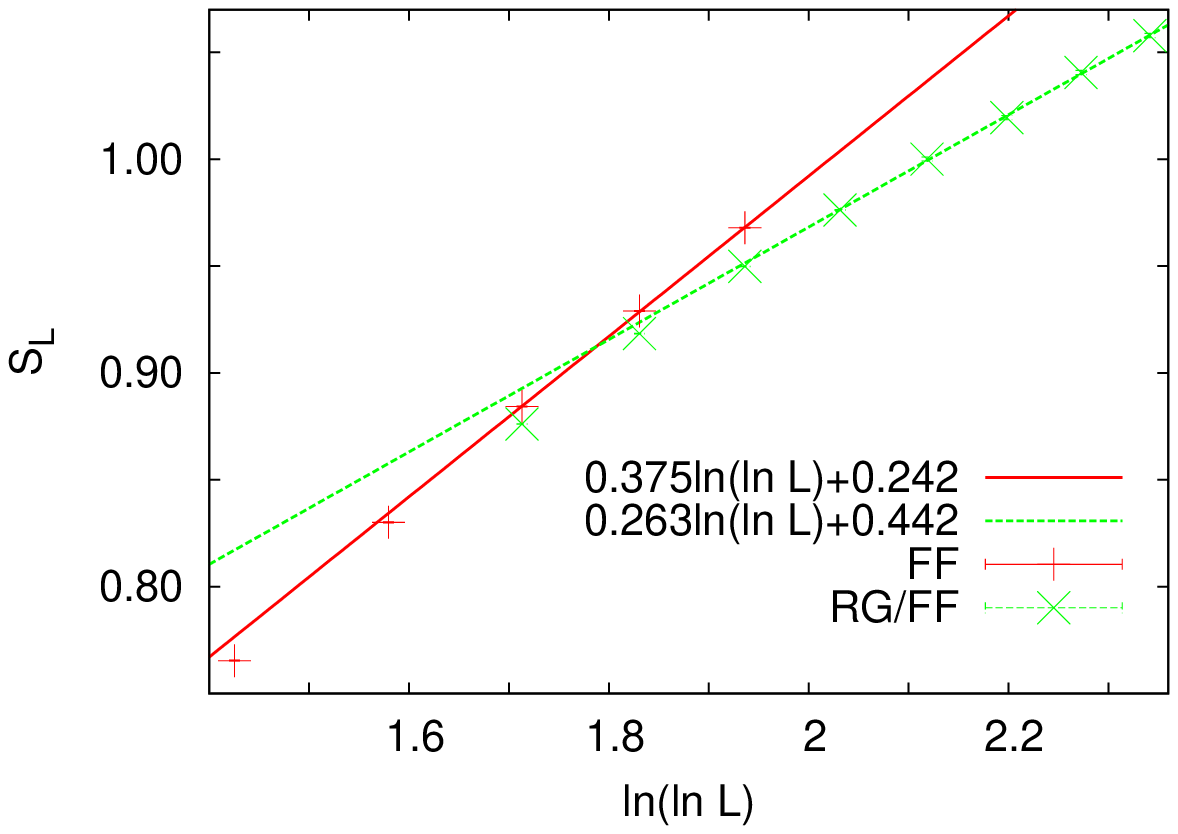}
\includegraphics[width=9cm]{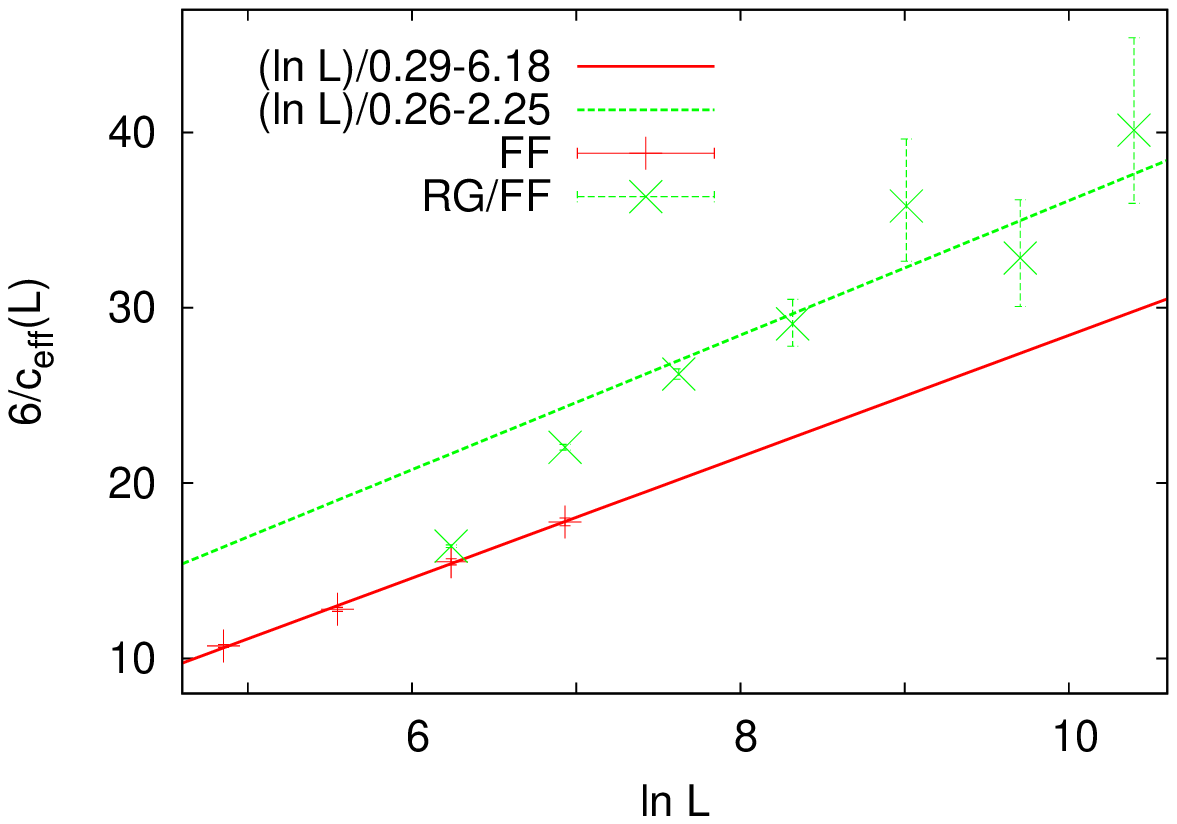}
\caption{\label{pr_S} (Color online) Top: size-dependence of the entanglement entropy in the composite AF XX-chain with a sharp interface obtained by free-fermion (FF) and the mixed (FF/RG) methods. Bottom: size dependence of the
inverse of the effective central charges.  
}
\end{figure}

In this section, we consider a non-random subsystem ${\cal B}$, with couplings $J_i=1$, whereas the other part ${\cal A}$ is random, in which the couplings are taken from the interval $[1/2,3/2]$. The composite chain is open, i.e. $J_0$ is taken randomly and $J_L=0$. The entanglement entropy was calculated numerically by the free-fermion method, the total length of the chains were $2L=64,128,\dots,2048$ and the number of random samples were $5 \times 10^5$. Having the results for the extended defect with $\kappa>1/2$ in the previous section, we expect a double logarithmic scaling of the average entanglement entropy. Indeed, in Fig.\ref{pr_S}, the data points of the average entanglement entropy as a function of $\ln \ln L$ are fit well to a straight line. 

We have also calculated the size-dependent effective central charge through two point fits and the results are shown in the bottom of Fig.\ref{pr_S}. In agreement with the expected double-logarithmic size dependence of the entanglement entropy, the inverse of the size-dependent effective central charge grows as $\ln L$. The prefactor of the double-logarithmic size-dependence of the entropy has been estimated from the size-dependence of $c_{\rm eff}(L)$, giving $\gamma \approx 0.29(3)$. This value is close, although somewhat different from the estimate of the prefactor obtained for the extended defect with $\kappa>1/2$.

%%%%%%%%%%%%%%%%%%%%%%%%%%%%%%%%%%%%%%%%%%%%%%%%%%%%%%%%%%%%%%%%%%%%%%%%%
%%%%%%%%%%%%%%%%%%%%%%%%%%%%%%%%%%%%%%%%%%%%%%%%%%%%%%%%%%%%%%%%%%%%%%%%%%%%%%%%%%%%%%%%%%%%%%%%%%%%%%%%%%%%%%%%%%%%%%%%%%%%%%%%%%%%%%%%%%%

\begin{figure}[h!]
\includegraphics[width=9cm]{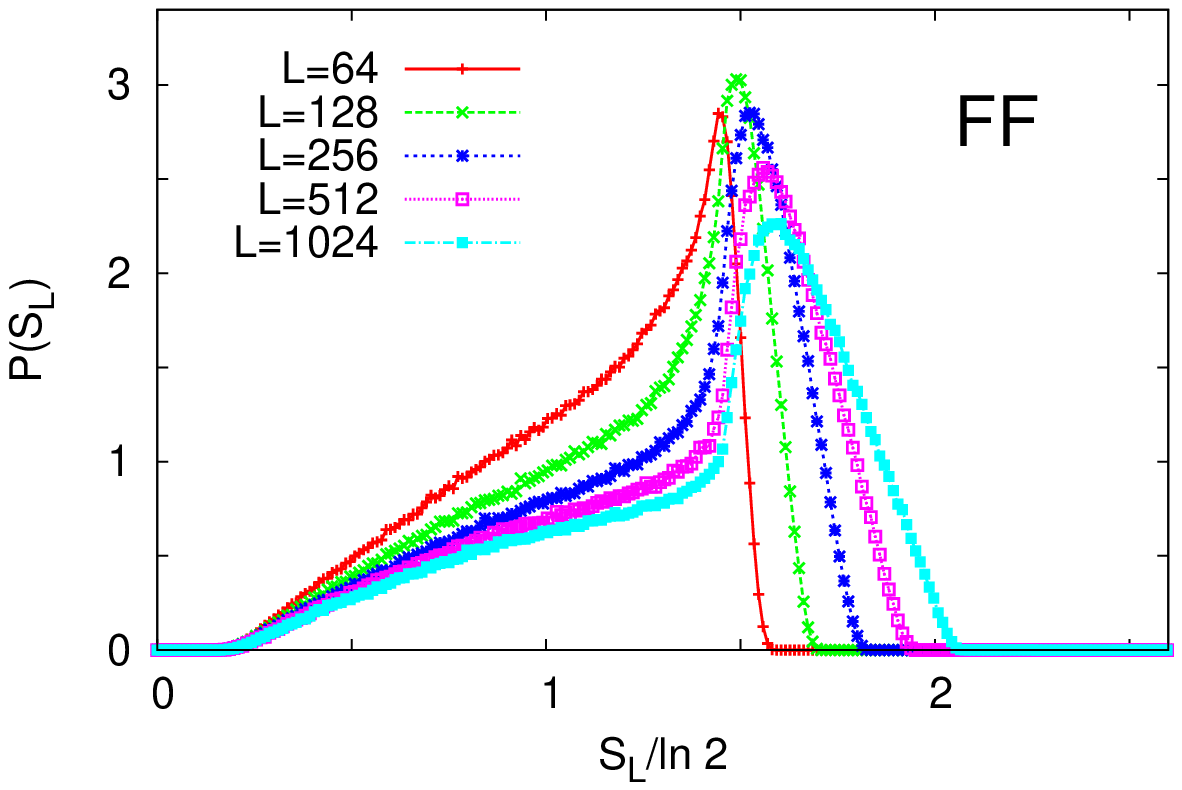}
\includegraphics[width=9cm]{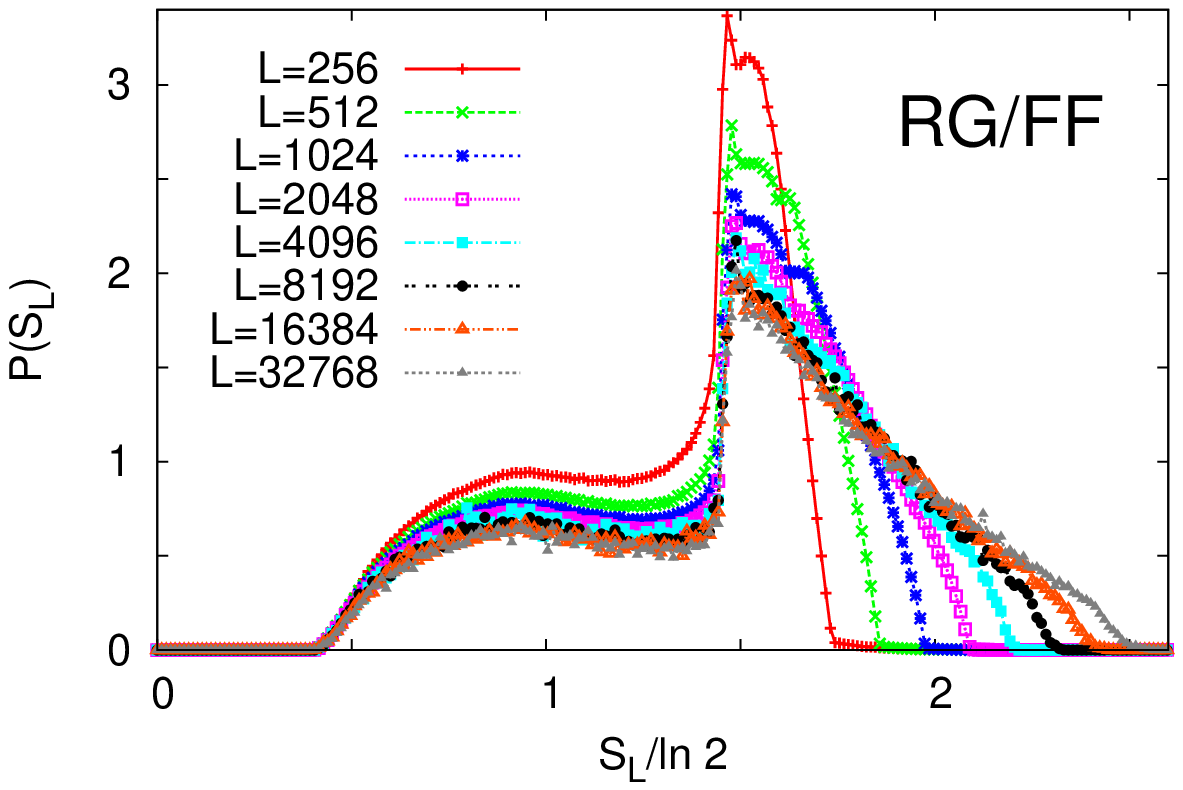}
\caption{\label{pr_hist} (Color online) Top: distribution of the entanglement entropy in the composite AF XX-chain with a sharp interface obtained by free-fermion (FF) methods. Bottom: the same calculated by the mixed (RG/FF) method.  
}
\end{figure}

The distributions of the entanglement entropy for different lengths are shown in Fig.\ref{pr_hist}. We have checked that the positions of the maxima of the distribution are shifted approximately as $\ln\ln L$ and at the same time the distributions broaden with approximately in the same scale. This type of scaling behavior is reflected in the size-dependence of the average entanglement entropy. We should mention, however, that the shape of the distributions has some variation with the size, thus even for the largest value of $L$ the distributions are still not in their asymptotic form.

%%%%%%%%%%%%%%%%%%%%%%%%%%%%%%%%%%%%%%%%%%%%%%%%%%%%%%%%%%%%%%%%%%%%%%%%%
%%%%%%%%%%%%%%%%%%%%%%%%%%%%%%%%%%%%%%%%%%%%%%%%%%%%%%%%%%%%%%%%%%%%%%%%%%%%%%%%%%%%%%%%%%%%%%%%%%%%%%%%%%%%%%%%%%%%%%%%%%%%%%%%%%%%%%%%%%%

In the following, we argue that the possible origin of the double-logarithmic scaling of the average entropy is due to the different types of ground states in the two separated (clean and random) subsystems. In the clean system in the ground state wave-function, the different spins are involved in the same collective form, whereas in the random AF XX-chain the ground state is a random singlet phase, each spin is strongly correlated with only one another spin. These random singlets are the building elements of the wave-function in the random side of the composite chain, and any spin which is involved in a random singlet pairing has negligible contribution to the entanglement in the composite system. Consequently, only the active spins, which are not involved in random singlet pairing, can contribute to the entanglement. 

To check this idea we have performed an approximate RG procedure in which the SDRG rules are used \textit{locally}. In the practical calculation in the random part of the chain, we checked all nearest neighbour bonds at the same time and all such bonds were transformed to a fixed singlet, if its strength was larger, than any of its two neighbours. Having calculated the renormalised values of the couplings over the singlets we repeated this procedure iteratively, and at the final stage, we arrived at a sequence of active sites. 
The idea of this approach is presented in more detail in Appendix \ref{appendix:rg}. It is shown there, that the number of active sites, $n_A(L)$, in a composite chain of length $L$, has an asymptotic logarithmic size-dependence. Considering the effective couplings between remaining active sites these are very week, their logarithm scales as: $\ln {\tilde J}_{i,j} \sim |i-j|^{1/2}$, $i$ and $j$ are the positions of the sites in the original, non-renormalised lattice. 

In the next step, we have joined together the renormalised random part of $n_A(L) \sim \ln L$ active sites and the clean part of length $L$ and calculated the entanglement entropy by the free-fermion method. For this composite system, we have used a special algorithm, which requires ${\cal O}(L^2)$ operations, and which is described in Appendix \ref{appendix_B}. Due to the use of the faster algorithm, we could go up to sizes $L=2^6$ and the average was performed over $10^5$ samples. The average entanglement entropy calculated in this way is presented in Fig.\ref{pr_S} as a function of $\ln\ln L$. The data points in Fig.\ref{pr_S} are in agreement with the double-logarithmic size-dependence, having a prefactor $\gamma \approx 0.26(6)$, which is somewhat smaller than the value found by the direct calculation in Fig.\ref{pr_S}. We have also calculated the distribution of the entanglement entropy by this approximate method. These are shown in the bottom part of Fig.\ref{pr_hist}. The shape of the distributions is similar to that obtained through the direct use of the free-fermion method.

%%%%%%%%%%%%%%%%%%%%%%%%%%%%%%%%%%%%%%%%%%%%%%%%%%%%%%%%%%%%%%%%%%%%%%%%%%%
\section{Discussion}
\label{sec:discussion}
In this paper, we have studied the entanglement properties of composite quantum spin chains. In particular, we have investigated the entanglement between a random and a clean part of an AF XX-chain. This problem can be equivalently formulated in terms of critical transverse-field Ising chains. The interface between the two parts was either sharp, or was realized by an extended interface defect, so that the strength of disorder in the clean part decays from the contact points asymptotically as a power with the distance: $\sim l^{-\kappa}$. According to a scaling theory, the extended defect is an irrelevant perturbation for $\kappa>1/2$, which is in agreement with a direct calculation of the local order parameter at the edge of the defect. The average entanglement entropy for a sharp interface, as well as for an extended defect with $\kappa>1/2$ was found to show a double-logarithmic size dependence: ${\cal S}_L \simeq \gamma \ln \ln L$. According to the numerical results, the variations of the prefactor are small, pointing toward a possible universal value. 

The double-logarithmic scaling of the entanglement entropy is in contrast with the logarithmic dependence which is seen in the (in a broader sense) homogeneous (random or clean) models. This can be understood by noting that in the random part most of the spins form random singlet pairs and therefore do not contribute to the entanglement. In a composite chain the number of active spins, which are not involved in singlets scales as $n_A(L) \sim \ln L$. A disconnected random chain, having infinitely strong couplings at the boundaries, can similarly be renormalised with the local SDRG rules and  a set of active sites can be identified, the number of which scales logarithmically with the length, too. If two random chains are then joined together, so that the infinite boundary couplings are replaced with the real ones, one can continue the renormalization process with the result, that the active spins at the two parts form singlets with each other with considerable probability. This results in a $\ln L$ scaling of the average entanglement entropy between the two random parts.\cite{refael,huse} If, however, the random part with $n_A(L)$ active spins is joined to a clean part, then these active spins have no preferential counterparts in the clean part. We can consider approximately the joint system as a composite of $n_A(L)$ and $L$ active spins, having an entanglement entropy which scales as $\ln n_A(L)$. This reasoning is in agreement with the numerical findings. We have also studied the entanglement entropy in the composite system having an extended defect with $\kappa<1/2$, when the perturbation is relevant. In this case the entropy is found to be proportional to that in the homogeneously random one, having the relation ${\cal S}(\kappa)=(1-2\kappa){\cal S}(\kappa=0)$.

The results obtained in this paper for the AF XX-chain are valid for a class of another quantum spin chains which are composed from a random and a clean part. As described in Sec.\ref{sec:model} the results can be directly transferred to the transverse-field Ising chain due to an exact mapping between the two free-fermion systems. However, the double-logarithmic scaling of the entropy is expected to hold for another quantum spin chains, too, which cannot be mapped to free fermions. The first example is the random AF XXX-chain, for which the random part can be renormalized to a set of random singlets\cite{im,fisherxx} and the number of remaining active sites scales as $n_A(L) \sim \ln L$, like in the random AF XX-chain. Considering the effect of an extended interface defect in the random AF XXX-chain, we have some differences. According to the Harris-like criterion in Sec.\ref{sec:harris}, the extended defect is an irrelevant perturbation for $\kappa>1$, since the correlation-length critical exponent in the clean AF XXX-chain is $\nu=2/3$\cite{xxx}. In this regime, the entanglement entropy is expected to scale in a double-logarithmic fashion, too. On the other hand, the SDRG procedure for this model predicts infinite disorder scaling in the regime $\kappa<1/2$. Here, the same type of relation is expected to work for the entanglement entropy of the composite system as given in Eq.(\ref{S_kappa}). In the intermediate region, $1/2 \le \kappa<1$, a new type of scaling of the entanglement entropy is expected.

Another possible examples are the quantum $Q$-state Potts chains, which are generalizations of the transverse-field Ising chain ($Q=2$). The random quantum $Q$-state Potts chains have the same type of infinite-disorder fixed points for any value of $Q>2$\cite{im,potts}. The clean version of the chains has a quantum critical point for $Q\le 4$,
with a correlation-length exponent which varies with $Q$\cite{baxter}. In this regime, we expect double-logarithmic scaling of the entropy for a sharp interface, whereas, for an extended interface defect, the same scaling scenario is expected to hold for different values of $\kappa$, as for the AF XXX-chain. Only the limiting value of $\kappa$ for irrelevant perturbations will depend on the value of $Q$. For $Q>4$, the quantum phase transition in the clean Potts chain is of first order and the entanglement entropy at the transition point shows a jump\cite{lajko_igloi}, instead of a logarithmic singularity. It could be an interesting question, whether, in the composite system, the jump at the transition point survives, or the entropy becomes continuous due to the disordered part.

\begin{acknowledgments}
This work is dedicated to John Cardy on the occasion of his 70th birthday.
Our research was supported by the Hungarian Scientific Research Fund under Grants
No. K109577 and No. K115959. 
\end{acknowledgments}

\appendix
\section{Renormalization of a composite system by the rule of local maxima}
\label{appendix:rg}

Considering a composite system consisting of a homogeneous part $\mathcal{A}$ and an identically distributed part $\mathcal{B}$, one expects that the ground state in the latter part is essentially a random singlet state. We will argue that this is true, indeed, apart from a small number of spins that remain unpaired. 
Obviously, the SDRG method in its original formulation, which reduces gradually the global energy scale, cannot be applied to the composite system, as the homogeneous part remains intact and the procedure stops at the scale $\Omega=J_{0}$. 
Therefore we need an alternative implementation of the idea of reducing the number of degrees of freedom. 
Let us consider a pair of spins with a coupling $J_2$, which is greater than the neighbouring couplings $J_1$ and $J_2$. Then, although $J_2$ is not necessarily the largest coupling in the whole system, the state of the pair has a high overlap with a singlet state, irrespective of the magnitude of couplings in the rest of the chain.   
We can therefore apply the elementary reduction step of SDRG method for any local maxima of the sequence of couplings, i.e. for those fulfilling $J_2>J_1,J_3$. 
It is easy to see that, in case of a finite chain with identically distributed disorder, the elimination of local maxima in an arbitrary order down to two spins results in the same singlet pair structure and effective couplings as those obtained by the traditional implementation of the SDRG procedure.   

Applying the elimination of local maxima to part $\mathcal{A}$ of a composite system, in general, not all spins are arranged into pairs, but the procedure ends up with a set of spins, where the sequence of couplings connecting them decreases monotonically away from the interface.  
The spins that are arranged in singlet pairs have essentially no contribution to the entanglement entropy. This is, however, not the case for spins that remain unpaired under the procedure, and these are entangled with part $\mathcal{B}$. According to numerical calculations by the free-fermion technique, such a subsystem with a monotonically decreasing sequence of couplings, behaves similarly to a homogeneous subsystem, i.e. the entanglement entropy increases logarithmically with the size of the subsystem, although with a prefactor smaller than that of a homogeneous part.   
The main question is the number of such unpaired spins in a finite subsystem of size $L$, as well as their locations.
Let us consider an XX chain with couplings $J_i$ (between sites $i$ and $i+1$) 
$J_i=1$ for $i\le 0$, and identically distributed random couplings for $i>0$. 
Subsystem $\mathcal{A}$ consists of sites $i>0$. 

\begin{figure}[h]
\includegraphics[width=6cm]{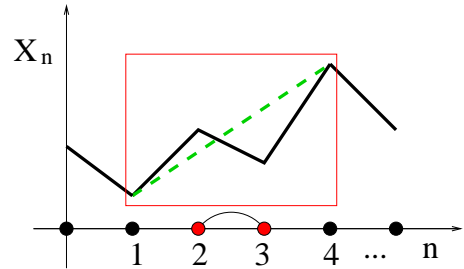} 
\caption{\label{zigzagrg} (Color online) 
Formation of a singlet by spins $2$ and $3$ and the corresponding change of the zig-zag path.   
}
\end{figure}

\begin{widetext}
%%%%%%%%%%%%%%%%%%%%%%%%%%%%%%%%%%%%%%%%%%%%%%%%%%%%%%%%%%%%%%%%%%%%%%%%%
\begin{figure}[h]
\includegraphics[width=17cm]{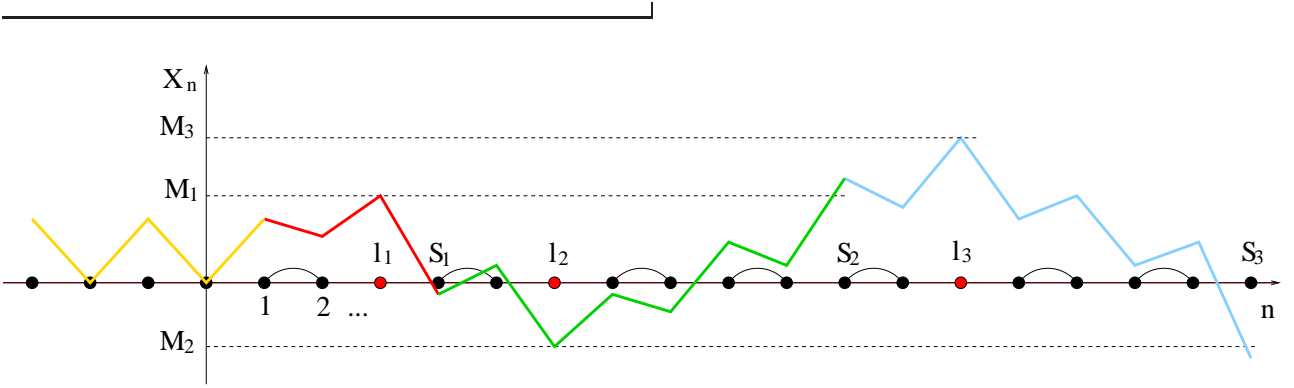}  
\caption{\label{seqsdrg} (Color online) A particular realization of a zig-zag path, which can be regarded as a process $X_n$ in discrete time. $S_N$, $N=1,2,3$ denote the times of record breaking events, while $l_N$ and $M_N$, $N=1,2,3$ denote the position and value of subsequent extrema, see the text. The former gives the positions of unpaired spins.  
}
\end{figure}
%%%%%%%%%%%%%%%%%%%%%%%%%%%%%%%%%%%%%%%%%%%%%%%%%%%%%%%%%%%%%%%%%%%%%%%%
\end{widetext}

As said above, the order of eliminations is unimportant, so one can start it at the interface and proceed in the disordered part. 
It is expedient to define a zig-zag path corresponding to a particular sequence of couplings by 
\beqn
X_0&=&0  \nonumber \\ 
X_n&=&\sum_{m=0}^{n-1}(-1)^{m}\beta_m,    \qquad (n>0)
\eeqn
where $\beta_m=\ln (\Omega_0/J_m)$.
The parts of the coupling sequence that fulfil the condition $\beta_2<\beta_1,\beta_3$ and are thus ready for decimation are represented by a three-segment path whose maximum and minimum are at its end points, see Fig. \ref{zigzagrg}.(This way of renormalization is analogous to that of the Sinai walk, see in Ref.\cite{im,sinai}.)  
The reduction step (formation of a singlet) then corresponds to the ``smoothing out'' of the three-segment part of the path.
%%%%%%%%%%%%%%%%%%%%%%%%%%%%%%%%%%%%%%%%%%%%%%%%%%%%%%%%%%%%%%%%%%%%%%%%%
%%%%%%%%%%%%%%%%%%%%%%%%%%%%%%%%%%%%%%%%%%%%%%%%%%%%%%%%%%%%%%%%%%%%%%%%
Note that $\{Z_n\}_{n=0}^{\infty}$ with $Z_n=X_{2n}$ can also be regarded as a symmetric random walk process in discrete time $n$. 
Let us carry out the SDRG procedure in terms of the zig-zag path, starting from the interface, see Fig. \ref{seqsdrg}. 
The path starts at $X_0=0$, and following and coarse-graining the path to site $n$, and one can easily see that spins arrange themselves into singlets up to this stage only if $X_m>0$ for $0<m\le n$. 
Assume that the first level crossing occurs at some $n=S_1$, (i.e. $X_{S_1}<0$, but $X_{m}>0$ for $0<m<S_1$). The path up to this point is coarse-grained to two segments, an ascending and a descending one, and the change in the latter is larger in magnitude than in the former. This means that the effective coupling corresponding to the second (descending) segment is weaker than that of the first segment, therefore the spin sitting at $n=l_1$, where the path has its maximum 
$M_1=X_{l_1}=\max_{0\le m<S_1}\{X_m\}$, will surely not form a singlet and remains unpaired.  

What is the probability distribution of the location $l_1$ of the first unpaired spin? First, we need the first-passage time $T_1=S_1-1$ at the origin of the process $X_n$ starting at $X_1=\beta_0$. As $\beta_0=O(1)$, and being interested in the large scale properties, the path can be approximately regarded as a Brownian bridge of length $T_1$. Its maximum position is typically at $O(T_1)$.  The probability of first return time to the origin of a symmetric walk is $p(t)\sim t^{-3/2}$, therefore the distribution of the distance of the first unpaired spin has the asymptotics 
\be 
p(l_1)\sim l_1^{-3/2}.
\ee
Note that the average of this distribution is infinite, $\overline{l_1}=\infty$. In finite disordered segments of size $L$, provided $L$ is even, the probability that no unpaired spins remain is $O(L^{-1/2})$.   
   
How far are the second and further unpaired spins? 
It is easy to see that, after the first level crossing event at time $n=S_1$, 
resetting time $n$ to zero and reflecting the increments $\Delta_n\equiv (-1)^n\beta_n$ of the path, $\Delta_n\to -\Delta_n$, we 
are up against with a similar first-passage time problem, however, with a new initial value. 
In this way, we have a sequence of processes, $\{Y^{(N)}_n\}_{n=0}^{L_N}$, $N=1,2,\dots$, each of which terminating on zero-level crossing at some time $L_N$. 
The first one ($N=1$) is $Y^{(1)}_n=X_n$, $n=0,1,\dots,L_1$, while the further ones are defined recursively as 
\beqn
Y^{(N)}_0&=&\mathcal{M}_{N-1}-Y^{(N-1)}_{L_{N-1}} \nonumber \\
Y^{(N)}_n&=&Y^{(N)}_0 + (-1)^{N-1}\sum_{m=1}^n\Delta_{S_{N-1}+m}, \qquad (n>0) 
\nonumber \\ 
\label{Yn}
\eeqn
Here, $S_N$, $N=1,2,\cdots$ denotes the successive first-passage times (without resetting of time), i.e.  $L_N=S_N-S_{N-1}$; 
and $\mathcal{M}_N=\max_{0\le n\le L_N}\{Y^{(N)}_n\}$ is the maximal value in the $N$th process.
We can see that the initial value for the $N$th process is the difference of the extremal values (the maximal and the final value) of the previous process. 
In each terminated process, an unpaired spin is produced at the maximum of the path, $l_N$, $Y^{(N)}_{l_N}=\mathcal{M}_N$, see Fig. \ref{seqsdrg}. 
Now, let us estimate the typical number of unpaired spins in part $\mathcal{A}$  of a finite length $L$ by a rough argument. 
In Eq. (\ref{Yn}), the final value $Y^{(N-1)}_{L_{N-1}}$ is $O(1)$, so 
the starting value for the $N$th process is approximately the maximum of the previous one, $Y^{(N)}_0\approx \mathcal{M}_{N-1}$. 
Starting from $Y^{(N)}_0$, the maximal value reached in the process is $a_NY^{(N)}_0$, and the typical value of the random variable $a_N$ is independent of $N$. So, $Y^{(N+1)}_0\approx a_NY^{(N)}_0$. 
In other words, $\overline{\ln Y^{(N)}_0}$, increases linearly with $N$.  
The logarithm of the typical first-passage time, $\ln L_{\rm typ}(N)\equiv\overline{[\ln(L_N)]}$, is thus also expected to increase proportionally to $N$. 
This means, that the typical time of processes, each containing one unpaired spin, increases exponentially with $N$, 
\be 
L_{\rm typ}(N)\sim e^{cN}.
\ee 
Equating the sum of typical times with the size $L$,
\be 
\sum_{N=1}^{\mathcal{N}}L_{\rm typ}(N)\sim L,
\ee
we obtain a logarithmic dependence of the typical number $\mathcal{N}$ of unpaired spins with $L$
\be 
\mathcal{N}\sim \ln L.
\ee
In other words, measuring the distance from the interface on a logarithmic scale, the unpaired spins are distributed uniformly. 

We can also establish a more direct connection between the locations of unpaired spins and certain extremal positions of the corresponding process $X_n$, as it is illustrated in Fig. \ref{seqsdrg}. 
Initially, the minimal value is zero, $M_0=0$. The process starts and terminates if $X_n<M_0$. If this event occurs (at some time $S_1$), we record the maximum value $M_1$ reached up to this time. Its position ($l_1$) gives, where the first unpaired spin is located. Then the process goes on until it breaks the record $M_1$ at some time $n=S_2$. Here, we look for the minimal value $M_2$ up to that time. Its position gives the position of the second unpaired spin. The process is then continued and terminated if the record $M_2$ is broken at some time $n=S_3$. The maximal value $M_3$ up to this time gives the new record and its position $l_3$ gives another unpaired spin. 
As can be seen, the locations of unpaired spins can be identified as subsequent times of record breakings of the minimum and the maximum in an alternating fashion as described above.

%\appendix
\section{Entanglement entropy between the renormalised random and clean chains}
\label{appendix_B}

The entanglement entropy between a random and a clean AF XX-chain with a sharp interface can be approximately calculated in two steps. As described in Sec.\ref{sec:disorder}, in the first step the random part of the chain is renormalised with the use of the local SDRG rules, by which we obtain a short chain of length $\ell \sim \ln L$ with inhomogeneous couplings. In the second step, the entanglement entropy between the renormalised part and the clean part is calculated by free-fermion methods. In fermionic systems, the entanglement entropy is calculated through the correlation matrix, which can be expressed with the components of the transformation matrix used in the diagonalization of the Hamiltonian through the Bogoliubov transformation\cite{vidal,peschel03,jin_korepin,peschel04,IJ07}. For the composite chain of lengths $\ell$ and $L$, one needs only the first $\ell$ components of this matrix, which can be efficiently calculated without the solution of the complete matrix\cite{note1}.

Let us denote the transformation matrix by $\mathbb{{\cal U}}$ and its $k$-th vector-component by $\mathbf{U}(k)$, which satisfies a tridiagonal eigenvalue problem: $\mathbb{{\cal T}}\mathbf{U}(k)=\epsilon(k) \mathbf{U}(k)$, see e.g. in Refs.\cite{ijr00,IJ07}.
Here, the non-vanishing elements of $\mathbb{{\cal T}}$ are $T(i,i+1)=T(i+1,i)=r_i$, for $i \le \ell$, which are obtained after performing the renormalization and $T(i,i+1)=T(i+1,i)=1$, for $\ell < i < L$, i.e. in the clean part. Now let us denote by $\mathbf{u}(k)$ a vector of length $\ell$, which contains the first $\ell$ components of $\mathbf{U}(k)$ and similarly by $\mathbf{w}(k)$ a vector of length $L$, having the last $L$ elements of $\mathbf{U}(k)$. Similarly, we write the matrix $\mathbb{{\cal T}}$ into block-matrix form and obtain for the eigenvalue equation:
\begin{equation}
\left[\begin{array}{c c}
 \mathbb{{\cal D}} & \mathbb{{\cal F}} \\ 
 \mathbb{{\cal F}}^T & \mathbb{{{\cal O}}}
\end{array}\right] 
\left[\begin{array}{c}
 \mathbf{u}(k) \\ 
\mathbf{w}(k) 
\end{array}\right]=
\epsilon(k)
\left[\begin{array}{c}
 \mathbf{u}(k) \\
\mathbf{w}(k) 
\end{array}\right] \;.
\label{block-matrix-form-shorthand}
\end{equation}
Here, the matrices $\mathbb{{\cal D}}$, $\mathbb{{{\cal O}}}$ and $\mathbb{{\cal F}}$ have the dimensions
$\ell \times \ell$, $L \times L$ and $\ell \times L$, respectively and $\mathbb{{\cal F}}^T$ is the transpose of $\mathbb{{\cal F}}$. We solve the eigenvalues of these equations by a standard LAPACK routine, which works in a computational time ${\cal O}((L+\ell)^2)$, but the eigenvectors are calculated separately. For this we notice, that the components of the vector $\mathbf{u}(k)$ satisfy the set of $\ell$ equations: 
\begin{equation}
 (\mathbb{{\cal D}}+\mathbb{{\cal F}}(\mathbb{{{\cal O}}} -\epsilon(k))^{-1})\mathbb{{\cal F}}^T) \mathbf{u}(k) = \epsilon(k) \mathbf{u}(k)\;,
 \label{reduced-block}
\end{equation}
and the inverse matrix, $(\mathbb{{{\cal O}}} -\epsilon(k))^{-1}$ is expressed in a closed form, see in Ref.\cite{matrix-inverse}. Having the eigenvalues at hand, we calculate the components of the vectors $\mathbf{u}(k)$ from Eq.(\ref{reduced-block}) numerically in ${\cal O}((L+\ell)\times \ell)$ time. The next step is to find the proper normalization of the complete vectors $\mathbf{U}(k)$. For this we should match the eigenvectors $\mathbf{w}(k)$, the components of which are expressed either as $ \sin(kl + \phi)$ or $\sinh(kl + \phi')$, with the numerical solutions of $\mathbf{u}(k)$ and calculate the norm of the resulting vector $\mathbf{U}(k)$. As mentioned earlier, to calculate the entanglement entropy we need only these properly normalized $\mathbf{u}(k)$ vectors.

%%%%%%%%%%%%%%%%%%%%%%%%%%%%%%%%%%%%%%%%%%%%%%%%%%%%%%%%%%%%%%%%%%%%%%%%%%%
%%%%%%%%%%%%%%%%%%%%%%%%%%%%%%%%%%%%%%%%%%%%%%%%%%%%%%%%%%%%%%%%%%%%%%%%%%%

\end{document}